\documentclass[aip,preprint]{revtex4-1}

\usepackage{graphicx}
\usepackage{dcolumn} 
\usepackage{bm}      
\usepackage{amsmath,amssymb,amsthm,amsbsy}

\draft 

\begin{document}


\title{On the relationship between the plateau modulus and the threshold frequency in peptide gels}

\author{L. G. Rizzi}

\affiliation{1.\,Depto.\,de\,F\'isica,\,Universidade\,Federal\,de\,Vi\c{c}osa,\,CEP:\,36570-000,\,Vi\c{c}osa-MG,\,Brazil.}

\date{\today}

\begin{abstract}
Relations between static and dynamic viscoelastic responses in gels can be very elucidating and may provide useful tools to study the behavior of bio-materials such as protein hydrogels.
An important example comes from the viscoelasticity of semisolid gel-like materials, which is characterized by two regimes: a low-frequency regime where the storage modulus $G^{\prime}(\omega)$ displays a constant value $G_{\text{eq}}$, and a high-frequency power-law stiffening regime, where $G^{\prime}(\omega) \sim \omega^{n}$.
Recently, by considering Monte Carlo simulations to study the formation of peptides networks, we found an intriguing and somewhat related power-law relationship between the plateau modulus and the threshold frequency, {\it i.e.} $G_{\text{eq}} \sim ( \omega^{*} )^{\Delta}$ with $\Delta = 2/3$. 
Here we present a simple theoretical approach to describe that relationship and test its validity by using experimental data from a $\beta$-lactoglobulin gel.
We show that our approach can be used even in the coarsening regime where the fractal model fails.
Remarkably, the very same exponent $\Delta$ is found to describe the experimental data.
\end{abstract}


\maketitle 


Scaling laws have been fruitful theoretical 
approaches to describe complex mechanical responses of gels~\cite{florybook,degennes}.
	Universal viscoelastic behaviors as those predicted by time-cure~\cite{martin1990macromol,martin1991review} and time-temperature~\cite{ferrybook} superposition principles are of particular interest due to their usefulness for some experimental data analysis methods~\cite{larsonbook}.

	In particular, there is a well-known viscoelastic behavior of gel-like  materials which is characterized by a plateau regime at frequencies lower than a threshold frequency $\omega^{*}$, {\it i.e.}~$G^{\prime}\approx G_{\text{eq}}$, and a power-law stiffening behavior given by $G^{\prime}(\omega) \sim \omega^{n}$ at frequencies higher than $\omega^{*}$.
	There are only a few theoretical approaches which attempts to explain the values observed for the exponent $n$; for instance, $n=1/2$ due to crosslink unbinding dynamics\cite{broedersz2010prl},
$n=2/3$ due to Rouse dynamics assuming a fractal model~\cite{martin1989PhysRevA}, 
and $n=3/4$ due to single semiflexible filament dynamics described by the wormlike chain model~\cite{gittes1998pre}.
	Experiments~\cite{raobook} however indicates that the values assumed by $n$ are significantly more widespread in a range between $0.1$ and $0.9$.
	Monte Carlo simulations on the formation of self-assembled peptides networks in the coarsening regime~\cite{rizzi2015prl} has also suggested a wide range of values~\cite{rizzi2016sm}.
	Intriguingly, the results of simulations unveiled a power-law relationship between the plateau modulus and the threshold frequency,
\begin{equation}
G_{\text{eq}} \sim ( \omega^{*} )^{\Delta} ~~~~.
\label{powerlaw_relationship}
\end{equation}
The simulations indicates that, unlike the scaling law for $n$, the power-law relationship given by Eq.~\ref{powerlaw_relationship} with exponent $\Delta=2/3$ for different network formation times is more robust and holds in spite of the interaction strengths between peptides~\cite{rizzi2016sm}.

	The analogy between the sol-gel transition and percolation has lead to important scaling relations~\cite{martin1990macromol,martin1991review} such as $G_{\text{eq}} \sim \varepsilon^z$ and $\tau_z \sim \varepsilon^{-y}$, where $\tau_z$ is ``the longest relaxation time'' and $\varepsilon=|\phi_c-\phi|/\phi_c$ is the ``distance'' to the transition, with $\phi$ being {\it e.g.} the bond probability.
	Indeed, for many gel-like materials which are very close to the sol-gel transition, not only those scaling relations are verified but also the relationship $G_{\text{eq}} \sim \tau_z^\Delta$, so that $n=z/y=\Delta$.
	Unfortunately, the relationship between $n$ and $\Delta$ seems to breakdown when the system is not too close to the transition, {\it i.e.}~at long cure times when the gel network is in the coarsening regime~\cite{furst2008prl,corrigan2009eurphysJE,donald2014langmuir}.

	Here we devise a theoretical approach to describe the relationship described by Eq.~\ref{powerlaw_relationship} for peptide gels in the coarsening regime.
	We include data from microrheology experiments in order to demonstrate both the validity of our approach and the agreement with the results obtained in our previous simulations.


	Because of the weak mechanical response of peptide networks, the characterization of the viscoelasticity in gels is done mainly by microrheology techniques such as particle tracking videomicroscopy and light scattering methods~\cite{waigh2005review}.
	By probing the mean squared displacement 
$\langle \Delta r^2 (t) \rangle$ of
nano-sized particles,
one can extract the complex shear modulus $G^{*}(\omega)$
using a generalized Stokes-Einstein relation~\cite{squires2010annrev,waigh2016review}.
	The most common theoretical expression that describes experimental data for the mean squared displacement (MSD) comes from the assumption that the gel network has a fractal structure~\cite{krall1997physA,krall1998prl}, and it is a result closely related to the percolation description of the system, which reads
\begin{equation}
\langle \Delta r^2 (t) \rangle =
\delta^2 \left[
1 - e^{-(t/\tau_p)^p}
\right] ~~~~,
\label{KW_fractal_model}
\end{equation}
where $p$ is the exponent of a power-law observed at time intervals $t$ much lower than the characteristic time interval $\tau_p$, and $\delta^2$ is the limiting (or plateau) value of 
$\langle \Delta r^2 \rangle$ when $t\gg \tau_p$.
	Such expression have been largely used to fit experimental data but in the most of cases the gels are in the coarsening regime, so subtle discrepancies which have not yet being fully appreciated might have impaired further developments of that theory.
	For example, because $\langle \Delta r^2 \rangle$ approaches $\delta^2$ very quickly as $t>\tau_p$, corrections have to be made to accommodate master curves fitting~\cite{romer2014epl}, and Eq.~\ref{KW_fractal_model} seems to not give a fully consistent behavior between the MSD
and the diffusion coefficient $D(t)$ measured in dynamical light scattering experiments~\cite{teixeira2007jphyschemB}.
	Also, the assumption that gelation involves the formation of self-similar structures in the coarsening regime might not be valid (as indicated by experiments on $\beta$-lactoglobulin gels~\cite{gosal2004biomacromol}).

	Our approach starts by considering that an expression very similar to Eq.~\ref{KW_fractal_model} can be derived from a completely different theory by assuming the diffusion of a probe particle with radius $a$ subjected to a potential $U$ due to the gel network~\cite{holek2014jphyschemB}.
	In this case, the time dependent position distribution function $f(\vec{r},t)$ can be determined by the following Fokker-Planck type of equation~\cite{grootmazurbook}:
\begin{equation}
\frac{\partial f}{\partial t} =
- \nabla \left[ 
k_B T \beta(t) \nabla f + 
\beta(t)f \nabla U
\right] ~~~,
\end{equation}
where $k_B$ is the Boltzmann's constant, $T$ is the absolute temperature, and $\beta(t)$ is a Onsager's coefficient~\cite{holek2014jphyschemB}.
	By assuming a harmonic potential, $U(r)= \kappa \, r^{\,2}/2$, one can obtain the following general expression for the mean squared displacement~\cite{holek2014jphyschemB}
\begin{equation}
\langle \Delta r^2 (t) \rangle =
\frac{d k_BT}{\kappa}
\left[
1 - \exp \left(
-2 \int \beta(t')dt'
\right)
\right] ~~,
\label{holek_general_MSD}
\end{equation}
where $d$ is the number of degrees of freedom of the random walk, {\it e.g.}~$d=2$ for passive tracking videomicroscopy.
	Hence, the mechanical properties of the gel can be estimated assuming a regime where the radius $a$ of the probe particle is larger than the mesh size $\xi$ of the gel network~\cite{squires2010annrev}.
	One can recover Eq.~\ref{KW_fractal_model} from~\ref{holek_general_MSD} simply by assuming $\beta(t) \propto (t/\tau_p)^{p-1}$, which corresponds to a gaussian but non-markovian random walk
\cite{holek2006jcp,holek2014jphyschemB}.

\begin{figure}[!b]
\centering
\includegraphics[width=0.53\textwidth]{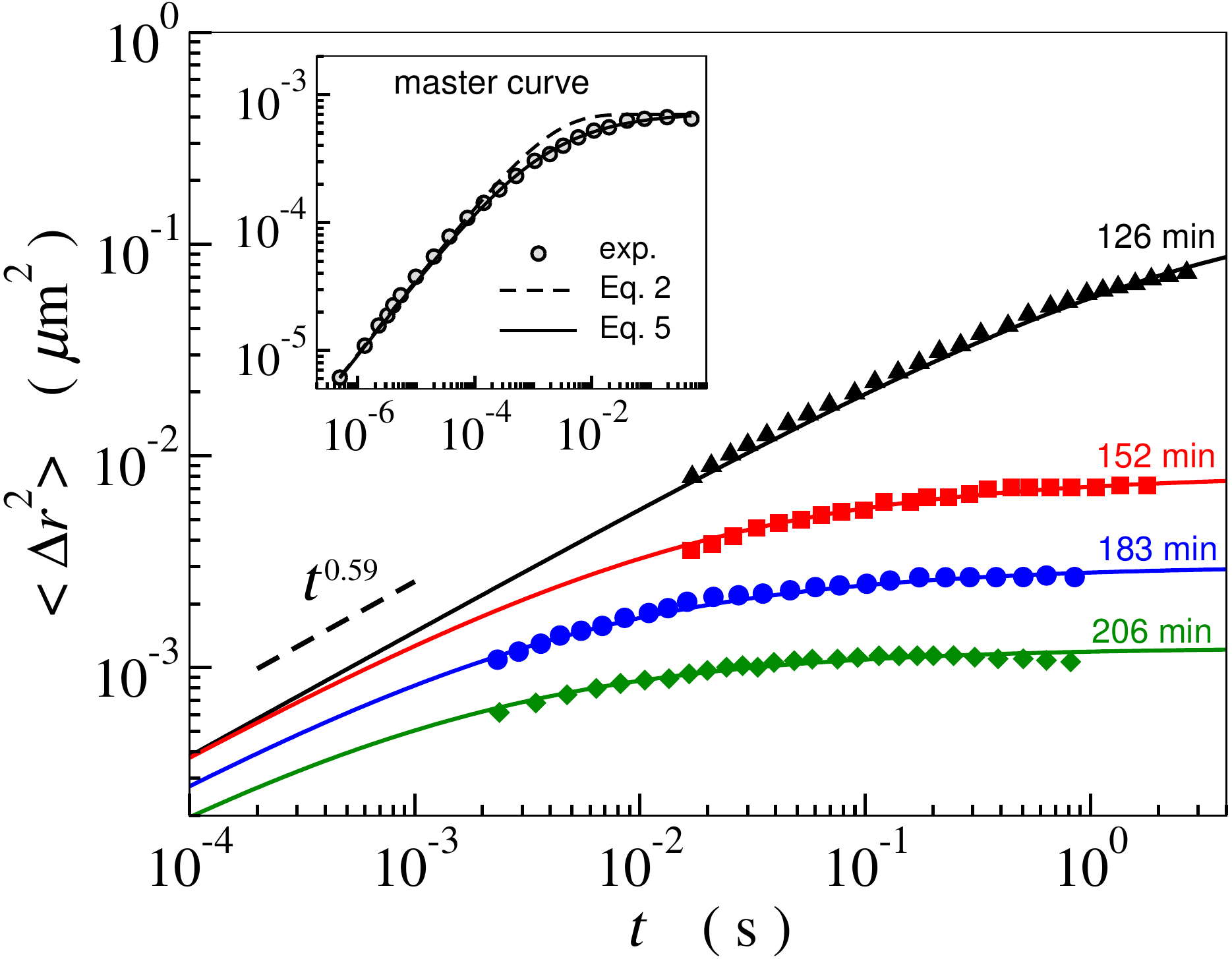}
\caption{Mean squared displacement $\langle \Delta r^2 \rangle$
of nano-sized polystyrene particles during the gelation of a
fibrillar $\beta$-lactoglobulin peptide gel.
The main panel shows different curves for different cure times $t_w=126$, $152$,
$183$, and $206$\,min. 
Filled symbols corresponds to experimental data extracted from Ref.~\cite{corrigan2009eurphysJE} while continuous lines denotes curves fitted using~Eq.~\ref{rizzi_solution} assuming fixed values for the exponents $n^{*}=0.59$ and $\alpha=0.746$ (which were obtained from the master curve), but different values of $\zeta^2$ and $\tau$.
Inset: experimental master curve (diamonds) and the curves fitted using Eq.~\ref{KW_fractal_model} (dashed line) and Eq.~\ref{rizzi_solution} (continuous line).}
\label{MSD}
\end{figure}

By adopting a slightly different function $\beta(t)$, one finds\footnote{A full description of the methodology used to obtain Eq.~\ref{rizzi_solution} will be described elsewhere.}
 that the MSD can be written as
$\langle \Delta r^2 (t) \rangle = \zeta^2 \left[ 1 - \exp{( - \alpha \ln \chi(t))}\right]$, 
with 
$\chi(t) = \left[  (  t/ \tau )^{n^{*}} + 1 \right]$,
so that the expression for the MSD becomes
\begin{equation}
\langle \Delta r^2 (t) \rangle 
= \zeta^2 
\left\{1-
\left[
  (  t/ \tau )^{n^{*}} + 1 
\right]^{-\alpha}
\right\}
 ~~,
\label{rizzi_solution}
\end{equation}
with the parameters $\zeta^2$ and $\tau$ being analogous to $\delta^2$ and $\tau_p$ in Eq.~\ref{KW_fractal_model}, respectively, but the exponent $p$ is substituted by two exponents, $n^{*}$ and $\alpha$.
	Our result provides a generalization of Eq.~\ref{KW_fractal_model} with the similar qualitative behavior
in the limiting cases,~{\it i.e.}~$\ln \chi(t) \sim (t/\tau)^{n^*}$ for $t \ll \tau$ and $\langle \Delta r^2 (t) \rangle \sim \zeta^2$ when $t \gg \tau$, as illustrated in Fig.~\ref{MSD}.

	Figure~\ref{MSD} includes data taken from Ref.~\cite{corrigan2009eurphysJE},
where the MSD curves were 
obtained by passive tracking microrheology using polystyrene probe particles with radius $a=500\,$nm.
	The experiments were performed with 3\% $\beta$-lactoglobulin (90\% pure) in water at pH=2 
and $k_B T=4.874$\,pN.nm ($T=80^{\text{o}}$C), and the different curves denotes the gel formation at different cure times $t_{w} > t_{\text{gel}}\approx 116$\,min.
	The inset of Fig.~\ref{MSD} includes the experimental master curve extracted from Ref.~\cite{corrigan2009eurphysJE} and a comparison between our result (Eq.~\ref{rizzi_solution}) and the fractal gel model (Eq.~\ref{KW_fractal_model}).
	We observe that if one chooses to fit the power-law behavior using the correct exponent, $p=0.59$, Eq.~\ref{KW_fractal_model}
did not work so well to fit the master curve (similar limitation has been already seen in {\it e.g.}~Ref.~\cite{romer2014epl}), while the expression~\ref{rizzi_solution} fits the whole master curve.
	From the master curve we have obtained the two exponents $n^{*}=0.59$ and $\alpha=0.746$, which were used to fit all data obtained at the different cure times $t_w$, as shown in the main panel of Fig.~\ref{MSD}.
	As $t_w$ increases, both limiting value of the MSD ($\zeta^2$) and the characteristic time ($\tau$) decreases.
	Such behavior can be seen as consequence of the coarsening regime~\cite{corrigan2009eurphysJE}, where the movement of nano-sized probe particles becomes more restrict as the peptide network evolves to a more rigid structure.



	In order to evaluated the viscoelasticity of the peptide network during its gelation, we consider a relation between mean squared displacement and compliance~\cite{xu1998rheol,wirtz2009annrev,squires2010annrev}, which is given by
\begin{equation}
J(t) = \frac{3 \pi a}{d k_BT} \langle \Delta r^{2}(t) \rangle ~~~~.
\label{compliance}
\end{equation}
	In principle, the complex viscoelastic modulus $G^{*}(\omega)=G^{\prime}(\omega) + i G^{\prime \prime}(\omega)$, with $G^{\prime}(\omega)$ and $G^{\prime \prime}(\omega)$ being the storage and the loss modulus, respectively,
can be obtained by considering
the Fourier transform
of the stress relaxation
modulus~\cite{squires2010annrev}.
	However, since one usually measures $\langle \Delta r^{2}(t) \rangle$ from microrheology
experiments, the viscoelastic modulus can be conveniently evaluated as~\cite{manlio2009pre}
\begin{equation}
G^{*}(\omega) = \frac{1}{i \omega \hat{J}(\omega)} ~~~~,
\end{equation}
where the Fourier transform $\hat{J}(\omega) = \mathcal{F}\{ J(t) \}$ is computed numerically from the time series of the compliance, Eq.~\ref{compliance},
using the direct method
proposed in Refs.~\cite{manlio2009pre,manlio2013newjphys}.

\begin{figure}[!b]
\centering
\includegraphics[width=0.53\textwidth]{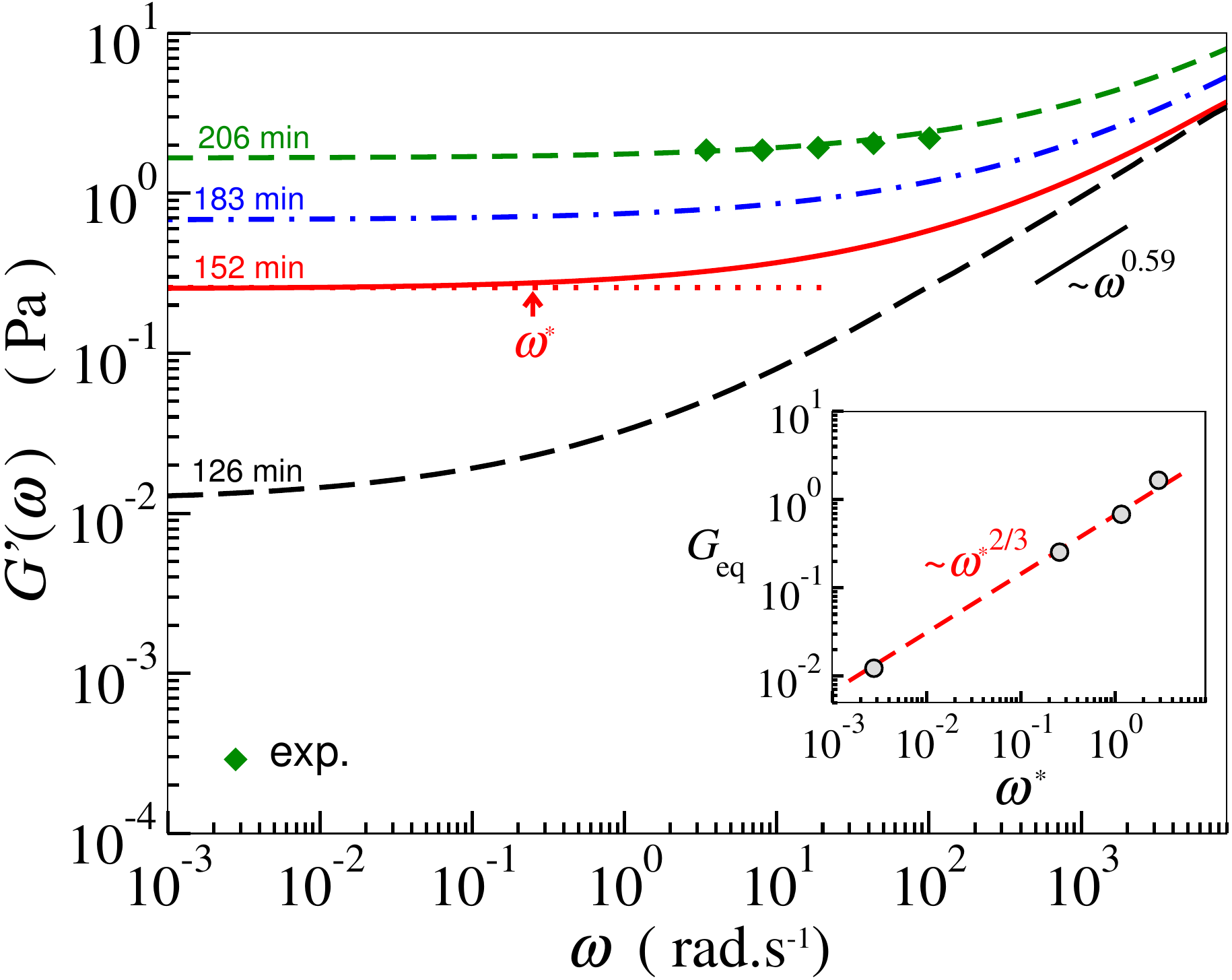}
\caption{Storage modulus $G^{\prime}(\omega)$ for the $\beta$-lactoglobulin peptide gel in the coarsening regime evaluated from the fitted curves, Eq.~\ref{rizzi_solution}, displayed in Fig.~\ref{MSD}.
Different curves denotes different cure times and the filled diamonds indicates experimental data obtained from Ref.~\cite{corrigan2009eurphysJE}.
The viscoelastic response of the gel is characterized by a threshold frequency $\omega^{*}$ which separates a plateau regime with $G^{\prime}\sim G_{\text{eq}}$ and a high frequency regime, where $G^{\prime}\sim \omega^{0.59}$.
Inset: 
power-law relationship between the plateau value $G_{\text{eq}}$ and the threshold frequency $\omega^{*}$ with exponent $\Delta=2/3$.}
\label{shearmodulus}
\end{figure}

	From a MSD curve described by Eq.~\ref{rizzi_solution} one should expect that for short times, $t/\tau \ll 1$, the mean squared displacement will behave as a power-law, {\it i.e.}~$\langle \Delta r^2 (t) \rangle \propto \zeta^2 (t/\tau)^{n^*}$, thus, at high frequencies ($\omega/\omega^* \gg 1$),
the storage
modulus should display a power-law behavior as well, {\it i.e.}~$G^{\prime}(\omega) \sim \omega^{n^*}$. 
	At low frequencies, which corresponds to $t/\tau \gg 1$ (or  $\omega/\omega^* \ll 1$),
Eq.~\ref{rizzi_solution} may lead to $\langle \Delta r^2 (t) \rangle \propto \zeta^2$, thus the storage
modulus $G^{\prime}(\omega)$ will be given by its plateau value~\cite{moschakis2013currop}, {\it i.e.}~$G_{\text{eq}} = d k_BT /3 \pi a \zeta^2$.

	Figure~\ref{shearmodulus} shows the storage modulus $G^{\prime}(\omega)$ evaluated from the MSD curves fitted with Eq.~\ref{rizzi_solution} 
at different cure times $t_w$.
	Accordingly, all the curves present the same qualitative semisolid gel-like viscoelastic behavior, that is, the storage modulus tends to a plateau value $G_{\text{eq}}$ at low frequencies and, at very high frequencies, it displays a power-law behavior $G^{\prime}(\omega) \sim \omega^{n^*}$.
	We define the value of the threshold frequency $\omega^{*}$ as
the frequency where $G^{\prime}$
deviates more than $10\%$ from the
plateau value $G_{\text{eq}}$.
	Our results for the viscoelastic modulus in Fig.~\ref{shearmodulus} confirms that both $G_{\text{eq}}$ and $\omega^{*}$ increases as the cure time $t_w$ increases, which means that the  gel is getting stiff as the structures of the peptide network coarsens.
	Indeed, as shown in the inset of Fig.~\ref{shearmodulus}, the plateau modulus $G_{\text{eq}}$ and the threshold frequency $\omega^*$ display a power-law relationship as given by Eq.~\ref{powerlaw_relationship}.
	Remarkably, the data extracted from Ref.~\cite{corrigan2009eurphysJE} yield the very same exponent $\Delta=2/3$ obtained in our simulations~\cite{rizzi2016sm}.
	Besides, our previous simulations~\cite{rizzi2015prl} have been shown to successfully describe the qualitative behavior of experiments~\cite{gosal2004biomacromol} on the evolution of the low-frequency storage modulus of $\beta$-lactoglobulin gels in the coarsening regime.
	Obviously, the agreement with the experiments does not ensures the universality of the value $\Delta=2/3$, and further experimental measurements are needed in order access the full extension of our findings.
 
	It is worth mentioning that the results in Fig.~\ref{shearmodulus} indicates that one might get only effective values for the exponent $n$ when probing the power-law behavior of $G^{\prime}(\omega)$ at frequencies just above the threshold frequency $\omega^{*}$.
	Although those results suggests that it might be difficult to measure the exponent $n^*$ for the most of cases, such somewhat slowly varying behavior 
of the storage modulus is consistent with the widespread values reported for $n$ in the literature~\cite{raobook}.


	In summary, we have confirmed that the plateau modulus $G_{\text{eq}}$ and the threshold frequency $\omega^*$ can be related by a power-law relationship given as in Eq.~\ref{powerlaw_relationship}.
	Of course, further experimental evidence is required to determined whether the exponent $\Delta$ obtained here is universal, or, if not, which specific features of the system could lead to that value.
	Importantly, we have introduced an alternative analytical expression to describe the mean squared displacement $\langle \Delta r^2(t)\rangle$, which is the most common experimental output obtained from microrheology techniques.
	Although Eq.~\ref{rizzi_solution} is quite general to be applied to any semisolid gel-like material, it might be valid only when the random walk of the probe particles are described by gaussian distributions~\cite{holek2014jphyschemB}, which seems to be the case of the experimental data considered here~\cite{corrigan2009eurphysJE}.
	Because our expression does not depend on any assumptions about the fractality of the peptide network, it can be used to describe gels which do not necessarily form self-similar structures.
	Thus, our approach should help one to access informations about the viscoelasticity of gels beyond the scaling laws derived from the fractal gel model~\cite{krall1997physA,krall1998prl}, and even in the coarsening regime, where the analogy between sol-gel transition and percolation~\cite{martin1990macromol,martin1991review} is not expected to be valid.

	Finally, we note that analytical expressions for the MSD should be specially important for data analysis in microrheology, {\it e.g}~to obtain suitable time-dependent diffusion coefficients from the dynamic light scattering data~\cite{teixeira2007jphyschemB}, and to ``tune'' sampling data in optical tweezers experiments~\cite{manlio2013newjphys}.
	Also, our approach might provide an alternative way to explore the time-temperature superposition principle~\cite{ferrybook}, which is a work that is still in progress.


The author acknowledge helpful discussions with
Stefan Auer, David Head, Manlio Tassieri, and Alvaro Teixeira.

\vspace{1.0cm}



%

\end{document}